\documentclass[a4size,journal]{IEEEtran}
\usepackage{amsmath,amsfonts}
\usepackage{algorithmic}
\usepackage{algorithm}
\usepackage{array}
\usepackage[caption=false,font=normalsize,labelfont=sf,textfont=sf]{subfig}
\usepackage{textcomp}
\usepackage{stfloats}
\usepackage{url}
\usepackage{verbatim}
\usepackage{graphicx}
\usepackage{layouts}
\usepackage{cite}
\usepackage{units}
\usepackage{todonotes}
\begin{document}

\title{On the Capacity of Future Lane-Free Urban Infrastructure}

\author{Patrick Malcolm, Klaus Bogenberger
\thanks{Manuscript received March 15, 2026}}

\markboth{Submitted to IEEE Transactions on Intelligent Transportation Systems}%
{Malcolm \& Bogenberger: On the Capacity of Future Lane-Free Urban Infrastructure}


\maketitle

\begin{abstract}
In this paper, the potential capacity and spatial efficiency of future autonomous lane-free traffic in urban environments are explored using a combination of analytical and simulation-based approaches.
For lane-free roadways, a simple analytical approach is employed, which shows not only that lane-free traffic offers a higher capacity than lane-based traffic for the same street width, but also that the relationship between capacity and street width is continuous under lane-free traffic.
To test the potential capacity and properties of lane-free signal-free intersections (automated intersection management), two approaches were simulated and compared, including a novel approach which we call OptWULF.
This approach uses a multi-agent conflict-based search approach with a low-level planner that uses a combination of optimization and simple window-based reservation.
With these simulations, we confirm the continuous relationship between capacity and street width for intersection scenarios.
We also show that OptWULF results in an even utilization of the entire drivable area of the street and intersection area.
Furthermore, we show that OptWULF is capable of handling asymmetric demand patterns without any substantial loss in capacity compared to symmetric demand patterns.
\end{abstract}


\section{Introduction}
\begin{figure}[!t]
\centering
\includegraphics[width=88.566mm]{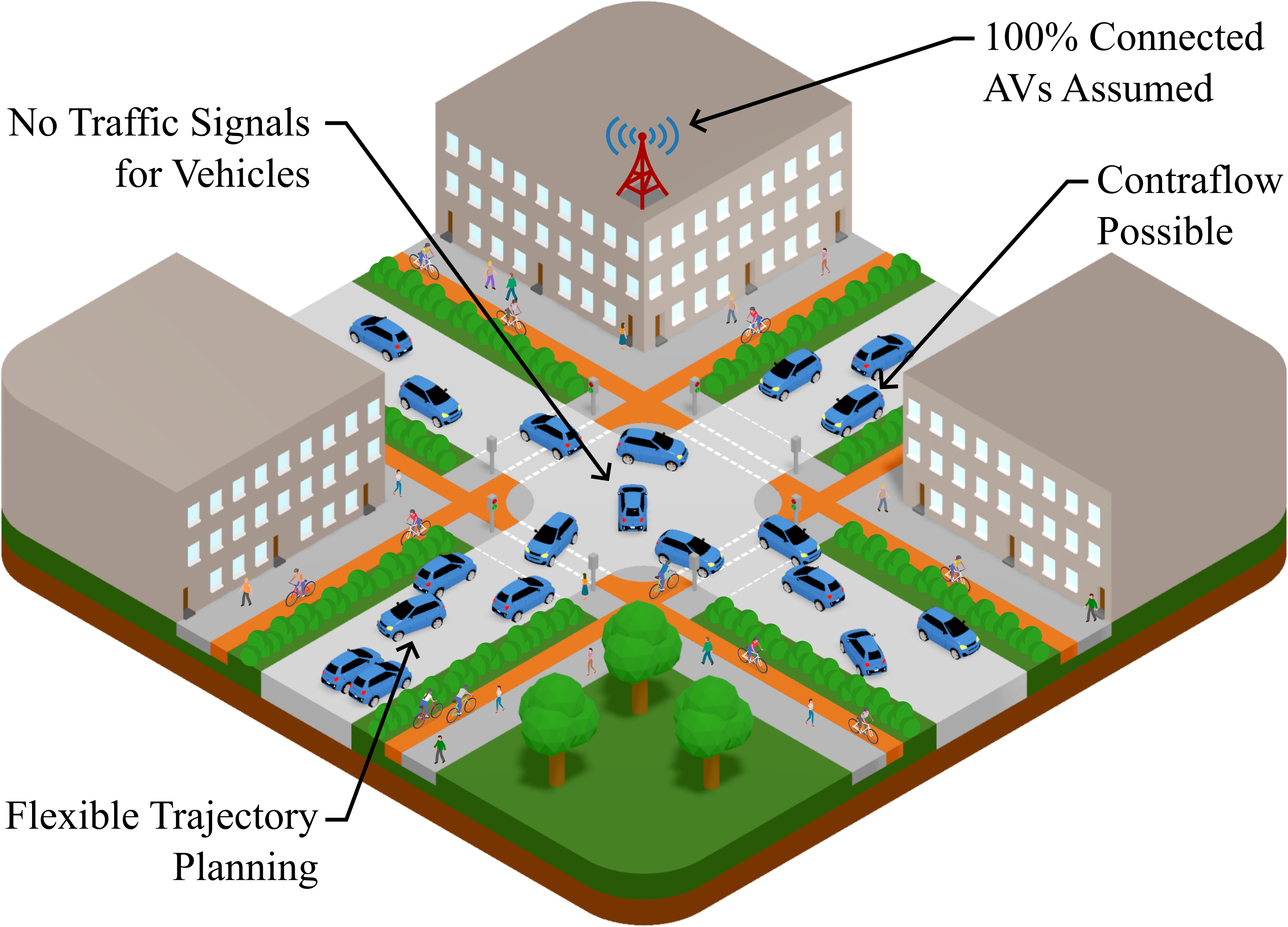}
\caption{Conceptual rendering of a lane-free intersection.}
\label{fig:lane-free-intersection-concept}
\end{figure}

%
\IEEEPARstart{I}{n} a future of widespread autonomous vehicles (AVs), it is possible to fundamentally change the ways in which traffic moves through our streets and intersections.
Two such concepts are automated intersection management (AIM) and lane-free traffic.
In the former, AVs navigate through an intersection without the need for conventional traffic signals, with each inter-AV conflict being resolved by the AIM control algorithm (which may be either centralized or decentralized).
In the latter, lane-free traffic, AVs are not required to adhere to pre-defined lanes but can instead freely choose their lateral position on the roadway at all times.
Since conventional lanes are substantially wider than a typical vehicle, and since AVs are likely to be able to drive with a reduced lateral gap between them compared to human-driven vehicles, lane-free traffic promises to provide an increased spatial efficiency (capacity per unit street width).
Increased flexibility in planning trajectories provides further increases in efficiency compared to lane-based traffic.
A conceptual rendering of a lane-free intersection is shown in Figure~\ref{fig:lane-free-intersection-concept}.

In this paper, the relationships between street width, capacity, and delay are explored in order to quantify the potential benefits and ramifications of the application of lane-free AIM to urban streets and intersections.
In order to estimate the maximum theoretical saturation flow of a street segment under lane-free traffic, an analytical approach is used in Section~\ref{sec:analytical-saturation-flow}.
For the investigation of the capacity and demand-delay relationship of urban intersections of different sizes, microscopic simulation is used.
Specifically, two AIM approaches are simulated.
The first of these is FERSTT \cite{Malcolm.2025.FERSTT}, which uses a rule-based spatiotemporal motion primitive tree search combined with a heuristic and penalty which encourages vehicles to keep as far right as possible unless it is disadvantageous to do so.
The second AIM is a novel approach called OptWULF, the workings of which are described in Section~\ref{sec:optwulf-description}. Selected results of simulations using FERSTT and OptWULF are provided in Section~\ref{sec:simulation-results}.

\section{State of the Art}
\label{sec:state-of-the-art}
Lane-based AIM has been an active area of research since the first publication on the topic \cite{dresner2004aim}. The goal of a lane-based AIM is to resolve conflicts between vehicle trajectories in the main intersection area.
To accomplish this, many approaches have been proposed, which can be categorized by the way in which they represent and resolve conflicts.
In a so-called ``safe pattern'' approach, the entire intersection area is considered as a single entity and only nonconflicting vehicles are allowed to occupy the intersection area at the same time. Tile-based approaches divide the the intersection area into a grid of discrete cells and ensure that no cell is occupied by more than one vehicle at a time. Conflict-point-~and conflict-region-based approaches are similar, but consider the precise location and geometry of each pair of vehicles. Lastly, vehicle-based approaches consider the vehicle trajectories directly without the need to represent the conflicts themselves. For a detailed review of these approach types and lane-based AIM in general, the reader is referred to \cite{zhong2021AIMlit}.
Initially, most research focused only on vehicle traffic, ignoring pedestrians and cyclists, though more recently, approaches incorporating all road user types have been proposed (e.g., \cite{niels2022dissertation}).

Lane-free traffic is a much newer, though very active area of research, with the first known academic work on the topic being done in 2015 \cite{johansenFlockingRoadTraffic2015}.
A large portion of the research on lane-free traffic to date has stemmed from the ``TrafficFluid'' concept proposed in \cite{papageorgiou2019trafficfluid}. Most of this research has focused on lane-free freeways and road segments rather than intersections. A comprehensive review of these works is provided in \cite{sekeran2022lanefreelit}.

Research into lane-free AIM is much more limited. Most literature on the topic focuses heavily on detailed vehicle kinematics modeling and control rather than the intersection control and traffic management aspects \cite{li2023SharingTrafficPriorities,amouzadiOptimalLaneFreeCrossing2023,ahmadi2024SignalFreePathFreeIntersection,huaNovelIntelligentIntersection2024,naderi2025LaneFreeSignalfreeIntersection,soltani2025CommunicationFreeDistributedModel}.
In \cite{stueger2023lanefreeaim}, an intersection-level lane-free control approach is proposed, but not thoroughly tested. The only known investigation into the capacity and demand-delay relationship of a lane-free AIM approach is \cite{Malcolm.2025.FERSTT}, in which the lane-free AIM FERSTT is used to simulate a symmetric demand pattern on a single small intersection of two streets of width \unit[6]{m}.
In this paper, we build on \cite{Malcolm.2025.FERSTT} by simulating a range of street widths as well as asymmetric demand patterns using not only FERSTT, but also a novel lane-free AIM approach which we call OptWULF.

\section{Analytical Derivation of Saturation Flow}
\label{sec:analytical-saturation-flow}
For a lane-based street, the saturation flow can be calculated analytically by assuming a certain minimum gross headway $\Bar{Z}$ between vehicles and number of lanes on the street $N_{lanes}$. Under these assumptions, the saturation flow is $q_{\max}=N_{lanes} \cdot \Bar{Z}^{-1}$. Assuming a typical (somewhat conservative) gross headway of $\Bar{Z}=\unit[2]{s}$ therefore results in a saturation flow of \unit[1800]{veh/h/lane}.
In order to apply this same analytical approach to a lane-free street, one needs to calculate an equivalent lane count $N_{lanes}'$ that represents the average number of vehicles which can drive side-by-side on a street of a given width. Given a statistical distribution of vehicle widths and an assumed lateral gap, the probability of $N$ vehicles fitting within the street width, denoted $p(N)$, can be calculated using Monte Carlo simulation.
The saturation flow of a lane-free street using this analytical approach is then:
\begin{equation}
\label{eqn:analytical-saturation-flow}
q_{\max} = \frac{N_{lanes}'}{\Bar{Z}} = \frac{1}{\Bar{Z}}\sum_{n=1}^\infty n \cdot p(N=n)
\end{equation}

\begin{figure}[!t]
\centering
\includegraphics[width=88.566mm]{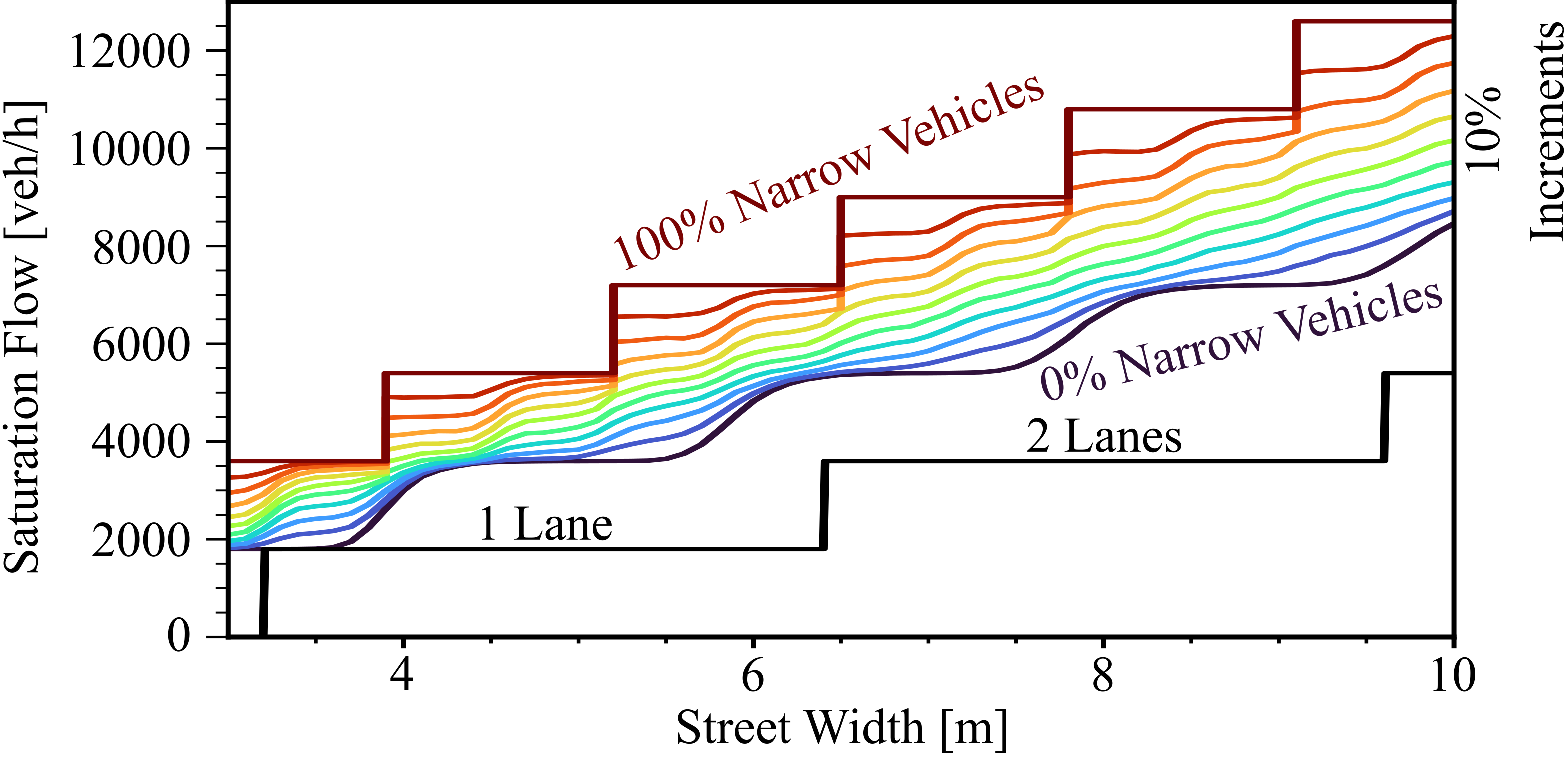}
\caption{Theoretical saturation flow of a lane-free street with various narrow vehicle penetration rates (colored lines, in 10\% increments) and a minimum lateral gap of \unit[0.1]{m} between vehicles, assuming a \unit[2]{s} gross headway. The lane-based saturation flow (assuming a lane width of \unit[3.2]{m}) is shown in black.}
\label{fig:saturation-flow-narrow-vehicles}
\end{figure}

Performing this procedure using a vehicle width distribution fitted to the InD data set \cite{Bock.2020.InD} ($w_{veh}\sim\mathcal{N}(\mu=1.87,~\sigma=0.14) ~\mathrm{with}~1.2<w_{veh}<2.8$~[m]) yields the results shown in Figure~\ref{fig:saturation-flow-narrow-vehicles}. This figure also shows the results under various penetration rates of narrow, single-passenger vehicles of width \unit[1.2]{m}, since under lane-free traffic, these could become more popular due to their increased trajectory planning flexibility.
Notable is the fact that even at a 10\% penetration rate, the relationship between saturation flow and street width is significantly smoothed out, and at 20\% it is nearly linear for street widths greater than \unit[6]{m}. As the narrow vehicle penetration rate approaches 100\%, the relationship again takes on a step function form (due to the assumed uniform width of all narrow vehicles). A continuous, near-linear, relationship between saturation flow and street width is interesting from a traffic engineering standpoint because it offers planners finer-grained control over the capacity of streets compared to lane-based traffic, under which capacity can only be increased or decreased in large discrete steps (by adding/removing an entire lane).
As illustrated, the key to this continuous relationship is sufficient variation in the widths of the vehicles.

\section{Description of OptWULF}
\label{sec:optwulf-description}
At a high level, the novel AIM approach described in this section uses a conflict-based search (CBS) in which the branching procedure attempts to resolve the new conflict in one of two ways: purely spatially and purely temporally. The approach can therefore be classified as a path-velocity decomposition approach. Spatial resolution of conflicts uses a simple window reservation process, whereas temporal resolution of conflicts involves solving an optimization problem to minimize the travel time through the intersection given a fixed spatial path. Because this approach uses a hybrid method which is both optimization-based and window reservation based, it is dubbed Optimization-based Window-planner for Urban Lane-Free Traffic (OptWULF).

As mentioned, OptWULF consists of three main components: the high-level CBS planner, the spatial conflict resolver, and the temporal optimization-based planner. After a description of the way in which OptWULF represents vehicle trajectories, each of these components is described in its own section below.

\subsection{Object Representation}
\begin{figure*}[!t]
\centering
\subfloat[]{\includegraphics[width=45mm]{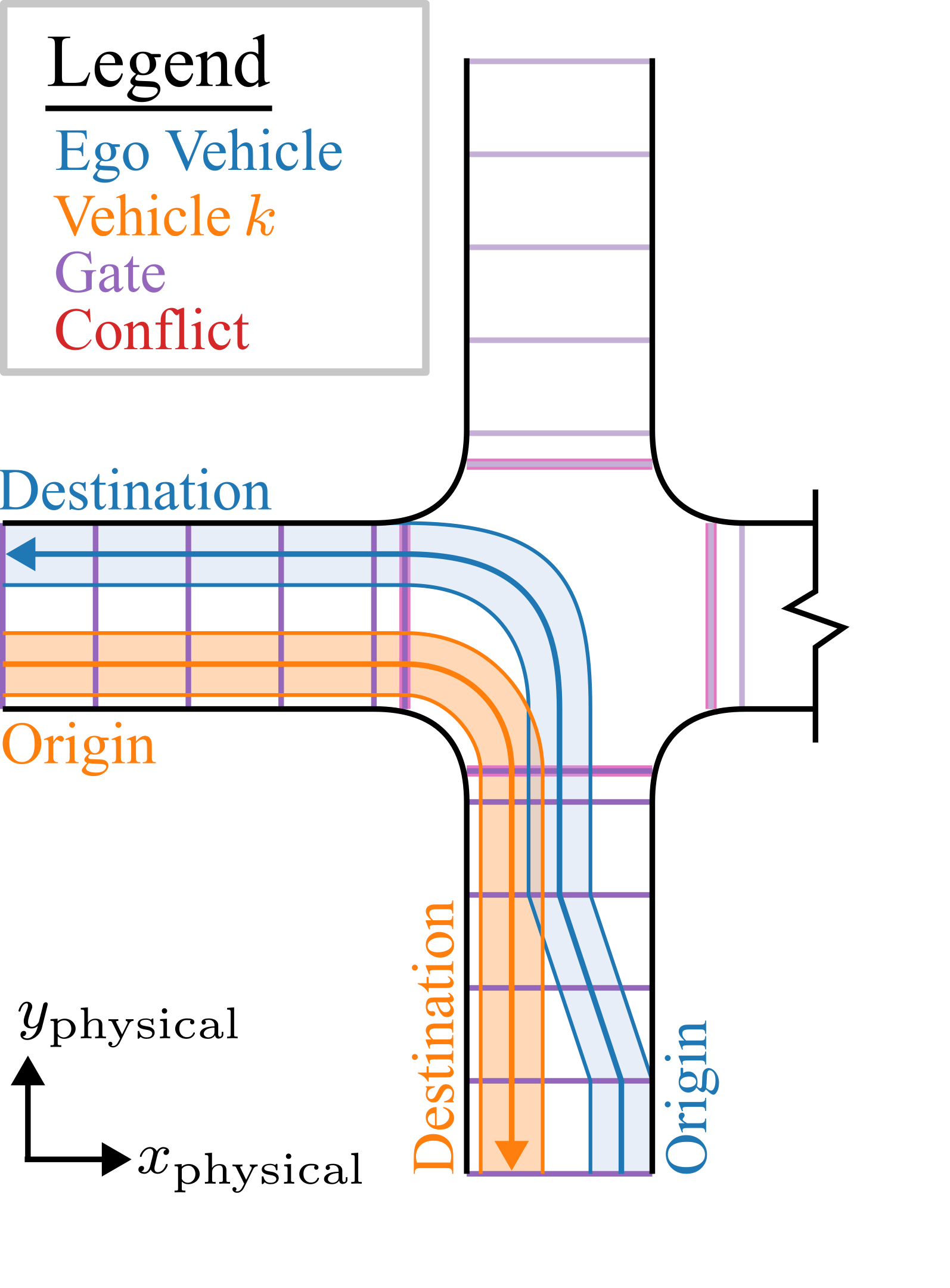}%
\label{fig:optwulf-object-plan-view}}
\hfill
\subfloat[]{\includegraphics[width=65mm]{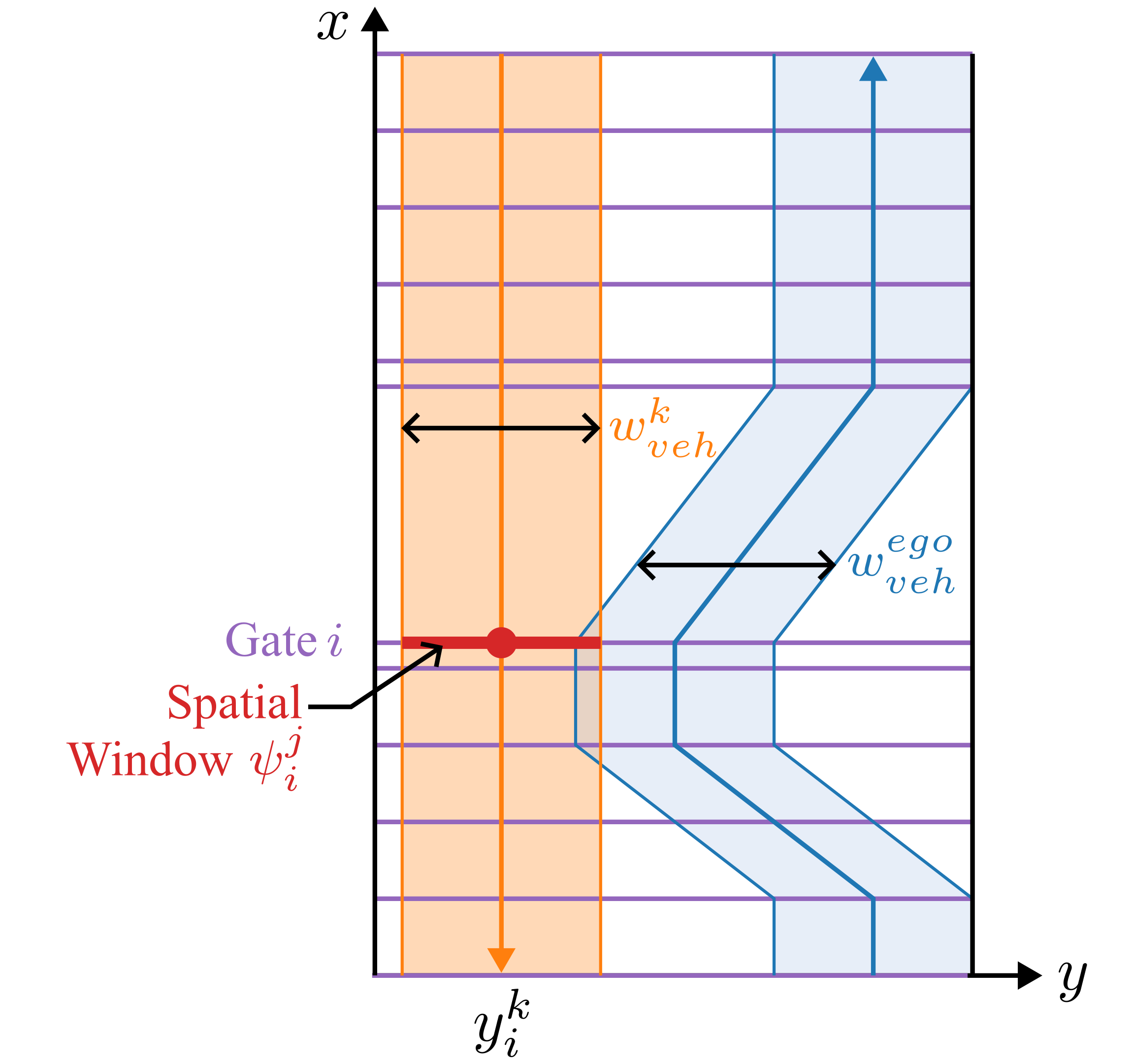}%
\label{fig:optwulf-object-spatial-view}}
\hfill
\subfloat[]{\includegraphics[width=71.35mm]{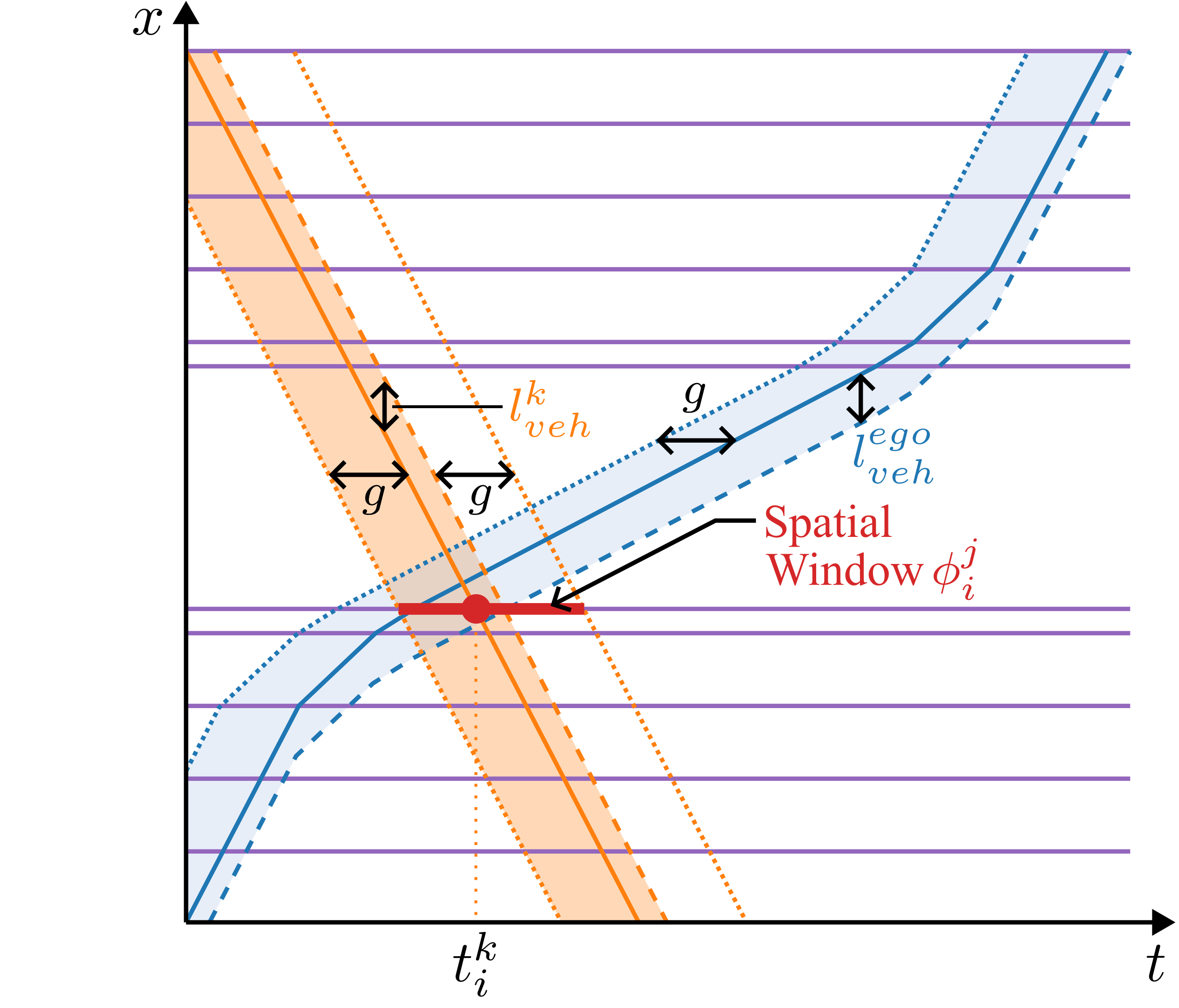}%
\label{fig:optwulf-object-temporal-view}}
\caption{Representation of intersection, trajectories, and conflicts in OptWULF. (a) Plan view. (b) Spatial alignment view. (c) Temporal view. Not to scale.}
\label{fig:optwulf-object-representation}
\end{figure*}

In OptWULF, the intersection and approach streets are represented as a series of gates through which vehicles must pass.
Each vehicle must maintain a minimum net time gap of $g$ (equivalent to post-encroachment time, PET) between it and any spatially conflicting vehicle trajectories.
A vehicle trajectory consists of a list of gates along with two values for each gate $i$: the associated lateral alignment $y_i$ indicating which part of the gate the vehicle crosses and the time $t_i$ at which the vehicle crosses the gate.
Together with the vehicle width $w_{veh}$ and length $l_{veh}$ and the minimum net time gap $g$, spatial and temporal envelopes for the vehicle trajectory can be drawn.
Between each gate, the trajectory is assumed to follow a straight line, and the distance between the gates is considered constant. The one exception to this is the segment passing through the intersection area itself. For turning vehicles passing through this segment, an optimal Dubins path~\cite{dubins1957} is dynamically calculated depending on the incoming and outgoing alignment. This means that a circular arc with the maximum possible radius is used, and any remaining distance to the incoming or outgoing point is covered by a linear segment. The length of this Dubins path is used as the segment length, and the radius $r_i$ is used to calculate the maximum velocity allowed for that segment.
Figure~\ref{fig:optwulf-object-representation} shows an example of the representation of two trajectories.

A conflict-free trajectory is one for which at each gate along its path, no other vehicle occupies the same gate with an overlapping spatial or temporal envelope (or ``window''). For the gates on the approach streets, the definition of these windows and any resulting conflicts is trivial. Special consideration must be given to the so-called ``intersection gates'' at the entrances to the main intersection area, however.

\begin{figure*}[!t]
\centering
\subfloat[]{\includegraphics[width=60.45mm]{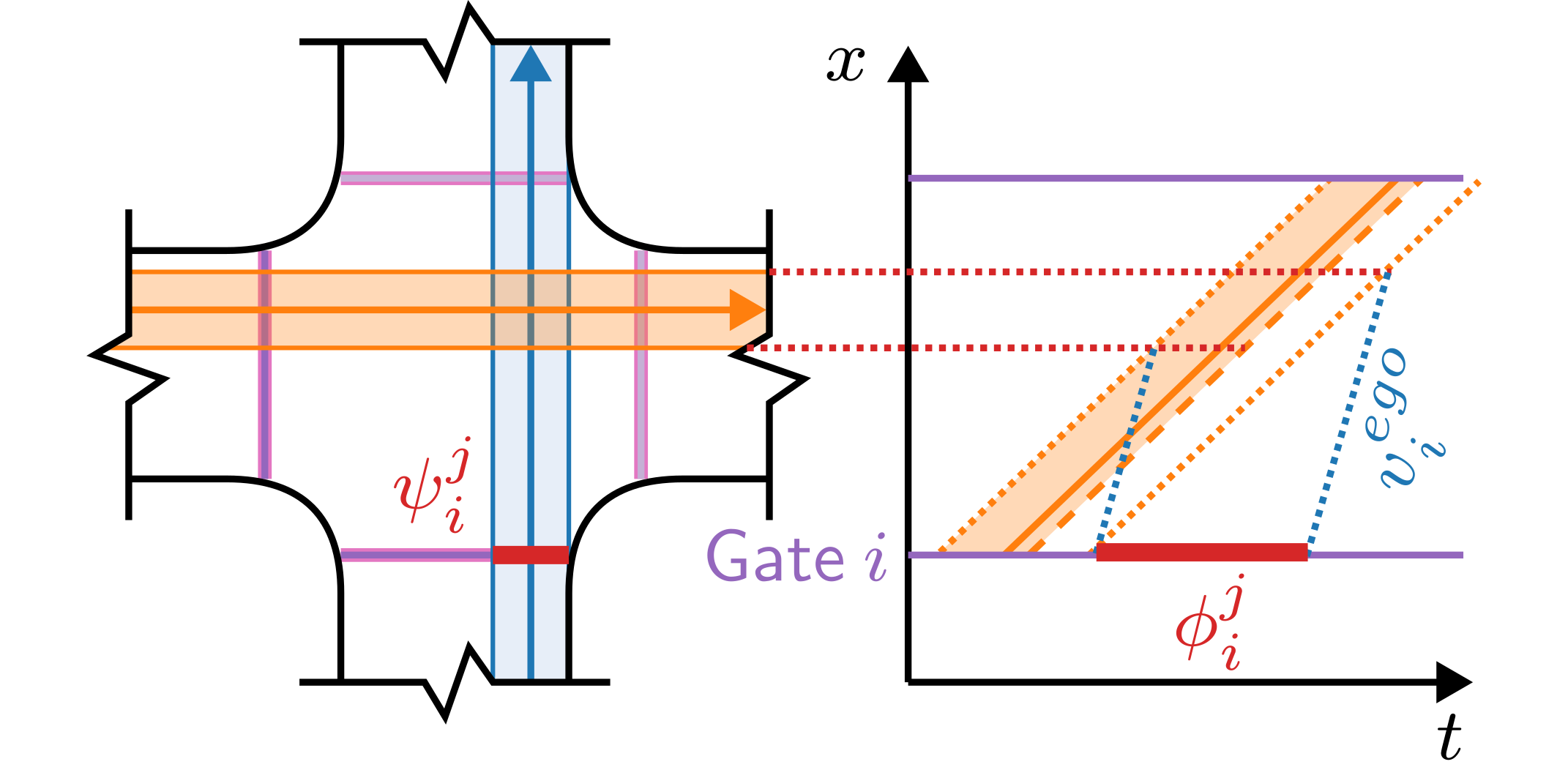}%
\label{fig:optwulf-conflict-perp}}
\hfil
\subfloat[]{\includegraphics[width=60.45mm]{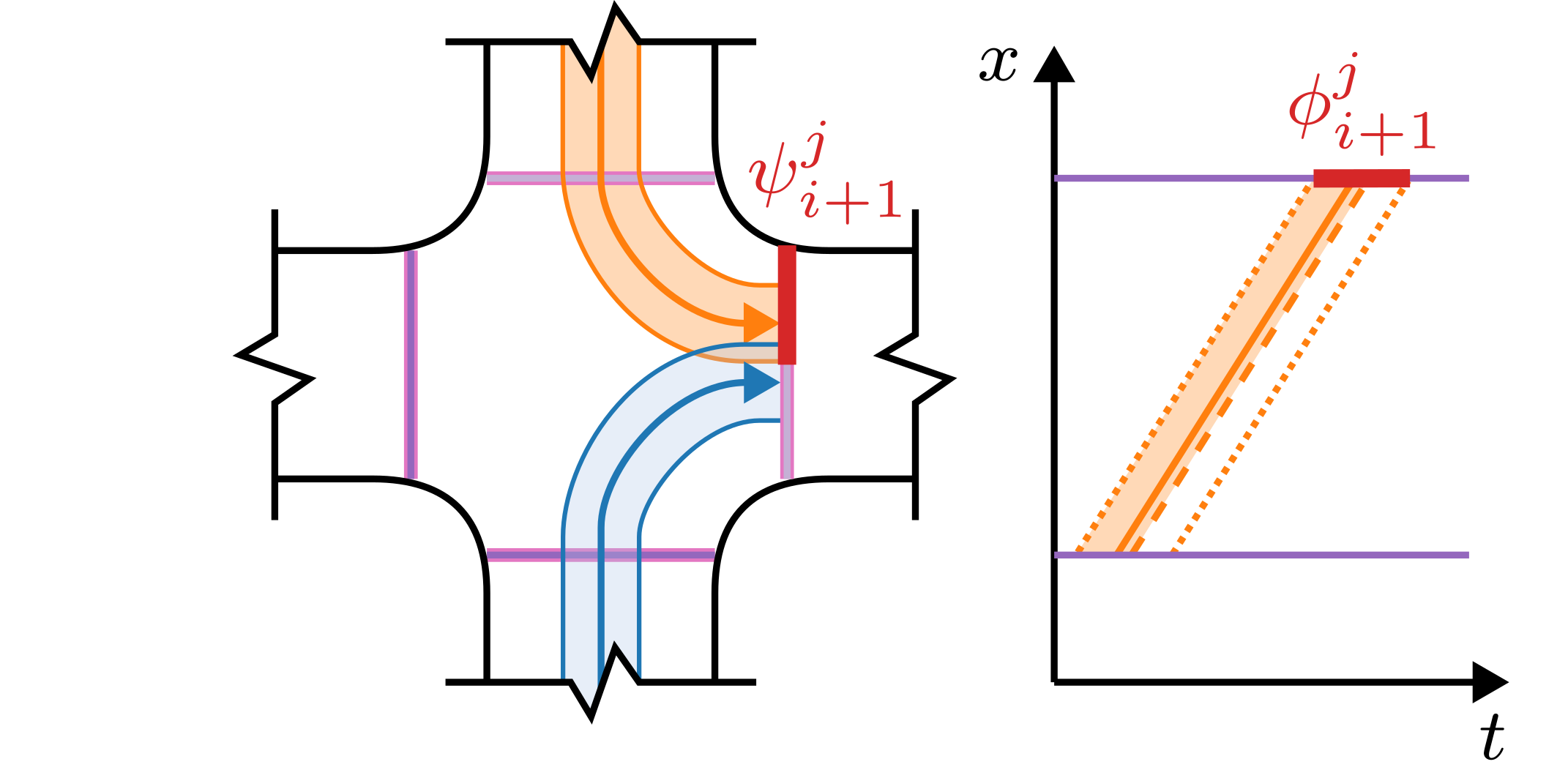}%
\label{fig:optwulf-conflict-merge}}
\hfil
\subfloat[]{\includegraphics[width=60.45mm]{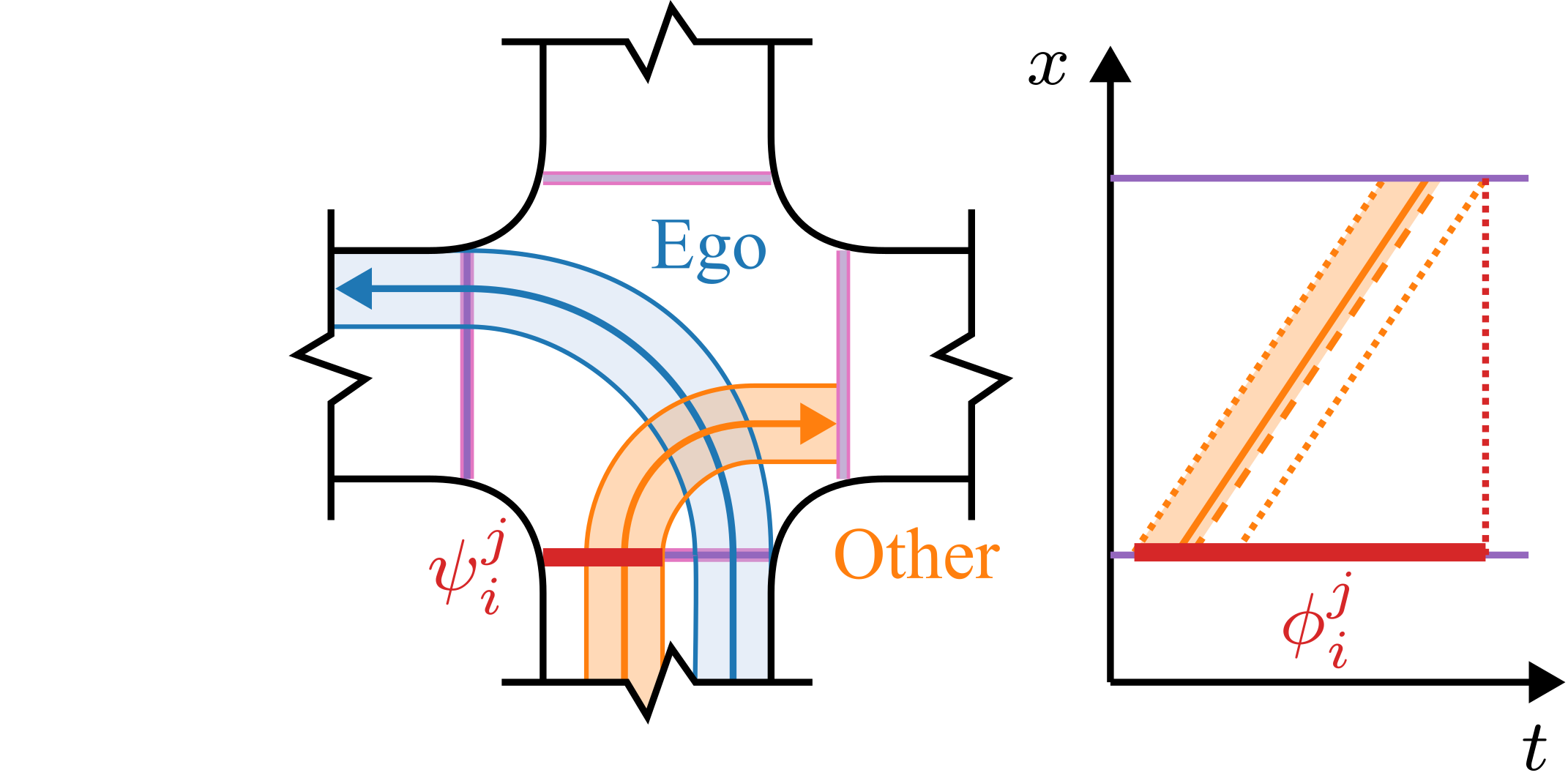}%
\label{fig:optwulf-conflict-share-origin}}
\caption{OptWULF special cases for intersection conflict definition. (a) Perpendicular conflicts. (b) Merging conflict. (c) Shared origin.}
\label{fig:optwulf-conflict-special-cases}
\end{figure*}

Every conflict $c_i^j$ for a gate $i$ consists of a spatial window $\psi_i^j$ representing the portion of the gate which is blocked and a time window $\phi_i^j$ representing the time range during which that part of the gate is blocked. Figure~\ref{fig:optwulf-object-representation} illustrates the definition of the windows $\psi_i^j$ and $\phi_i^j$ for the case of a conflict with an already-booked vehicle $k$ with lateral alignment $y_i^k$ and arrival time $t_i^k$ at gate $i$. These windows are defined mathematically as such:
\begin{align}
    \psi_i^j = & \left [ y_i^k - \frac{w_{veh}^k}{2} ~,~ y_i^k + \frac{w_{veh}^k}{2} \right ] \\
    \phi_i^j = & \left [ t_i^k-g ~,~ t_i^k+\frac{l_{veh}^k}{v_i^k}+g \right ] 
\end{align}

Minimum lateral gaps between vehicles are enforced by increasing $w_{veh}$ accordingly. Therefore, $w_{veh}$ represents the ``effective width'' of the vehicle rather than its actual physical width.

Traffic signal red phases can also be represented as conflicts in OptWULF. Their spatial window occupies the entire street width, and the temporal window is the time interval during which the signal is red for that cycle:
\begin{align}
    \psi_i^j = & \left [ 0~,~w_{street} \right ] \\
    \phi_i^j = & \left [ t_{red}^{start} ~,~ t_{red}^{end} \right ] 
\end{align}

Conflicts within the main intersection area need to be handled specially, since vehicles coming from different gates may still conflict with one another. The basic logic behind the detection of such a conflict between two vehicles is two treat their respective entry intersection gates as the same and their respective exit intersection gates as the same, with a few caveats. The conflict detection is broken down into several cases.

The first case, shown in Figure~\ref{fig:optwulf-conflict-perp}, is for perpendicular conflicts. For these conflicts, the actual conflict area within the intersection is calculated, and from this, the first vehicle's occupancy time in the conflict area (plus $g$) is projected onto the second vehicle's entry gate (using the second vehicle's speed) in a procedure analogous to the calculation of required intergreen times using clearance and entry times at conventional signalized intersections. In this case, the spatial window $\psi_i^j$ consists of the location of the ego vehicle at the entry gate.

The second case, shown in Figure~\ref{fig:optwulf-conflict-merge}, is for vehicles that enter from opposing directions and exit in the same direction (e.g.,\ one vehicle driving south to east and another driving north to east). In this case, the conflict is checked at the outgoing gate rather than the incoming gate. The temporal window $\phi_i^j$ is calculated in the same way as normal conflicts. The spatial window $\psi_i^j$ is the location of the other vehicle plus all locations on the opposite side from the ego vehicle (as in the first case above).

The third case, shown in Figure~\ref{fig:optwulf-conflict-share-origin}, applies when vehicles originate from the same origin. In this case, if their movements do not spatially cross, then no conflict is detected. If their movements do cross, then the temporal window $\phi_i^j$ of the conflict at the origin gate is extended by the other vehicle's time within the intersection. In other words, the conflict-resolving vehicle can not enter the intersection until the first vehicle has fully cleared the intersection (plus $g$). The spatial window $\psi_i^j$ for the conflict in this case is not only the location of the other vehicle, but also all locations on the other side of the vehicle (to prevent vehicles from ``jumping'' over one another laterally).

The fourth and final case is for all remaining conflicts. In this case, it is first checked whether there is any spatial overlap between the vehicles' spatial paths within the intersection area. If not, then no conflict is detected. If so, then both vehicles' entry gates are treated as the same, as are their exit gates, and the normal conflict detection procedure is applied. Similarly to the third case, this means that the second vehicle may not enter the intersection until $g$ seconds after the first has fully left it. The spatial window $\psi_i^j$ in this case is the location of the ego vehicle at the entry gate.

\subsection{CBS Planner}
The trajectory planner of OptWULF uses conflict-based search (CBS) \cite{sharon2015CBS}, which works as follows. When a new vehicle registers with the intersection controller, a trajectory is planned for it, ignoring all other already-booked vehicles' trajectories.
Then, the set of all trajectories (including that of the new vehicle and those of all previously booked vehicles) is checked for conflicts. If one or more conflicts is found between any pair of trajectories, then one of these conflicts is used to ``branch'', creating two independent ``sub-problems'', one in which the constraint applies to the first vehicle and one in which it applies to the second.
Each sub-problem $\mathfrak{P}$ consists of a list of all the vehicles which can be rerouted along with each of their trajectories and lists of constraints.
If one of the child sub-problems has an unresolved conflict, it will in turn result in further branching once it is visited, creating two new (sub-)sub-problems. The search space of the CBS procedure is the binary tree formed by these sub-problems (often called a ``constraint tree''), which it simultaneously generates and searches until it finds a sub-problem for which no vehicle trajectory conflicts with any other.

Each sub-problem $\mathfrak{P}$ has an associated cost, which is defined as the total travel time of all trajectories as planned in $\mathfrak{P}$. To ensure the optimality of the found solution, the as-yet-unvisited sub-problem with the lowest cost is always visited next.

When solving each sub-problem, the OptWULF planner chooses either a spatial or temporal conflict resolution depending on which has the lowest cost. For the spatial conflict resolution, the spatial conflict resolver is applied and a temporal component is determined for the new alignment using the temporal planner. For the temporal conflict resolution, the temporal planner is applied to the unchanged alignment. The entire high-level CBS procedure is described in Algorithm~\ref{alg:cbs}.

\begin{algorithm}[H]
\caption{OptWULF CBS}\label{alg:cbs}
\begin{algorithmic}
\STATE \textbf{create} initial problem $\mathfrak{P}_{\mathrm{init}}$ w/ already-booked trajectories
\STATE \textbf{add} unconstrained trajectory of new vehicle to $\mathfrak{P}_{\mathrm{init}}$
\STATE $\Vec{\mathfrak{P}} \gets \{\mathfrak{P}_{\mathrm{init}}\}$
\WHILE{$\Vec{\mathfrak{P}} \neq \emptyset$}
\STATE$\mathfrak{P} \gets$ \textbf{pop}$(\Vec{\mathfrak{P}})$ sub-problem w/ lowest cost 
\STATE$\mathfrak{P}_0,\mathfrak{P}_1 \gets$ \textsc{VISIT}$(\mathfrak{P})$
\STATE$\Vec{\mathfrak{P}} \gets \Vec{\mathfrak{P}} \cup \{ \mathfrak{P}_0,\mathfrak{P}_1 \}$
\ENDWHILE

\STATE
\STATE {\textsc{VISIT}}$(\mathfrak{P})$
\STATE\hspace{0.5cm}\textbf{check} for conflict $C$
\STATE \hspace{0.5cm}\textbf{if} C exists \textbf{then}
\STATE\hspace{1.0cm}$i,j \gets$ vehicles involved in $C$
\STATE\hspace{1.0cm}\textbf{copy} $\mathfrak{P}$ twice: $\mathfrak{P}_i$, $\mathfrak{P}_j$
\STATE\hspace{1.0cm}\textbf{add} constraint $C$ to trajectory $i$ in $\mathfrak{P}_i$
\STATE\hspace{1.0cm}\textbf{add} constraint $C$ to trajectory $j$ in $\mathfrak{P}_j$
\STATE\hspace{1.0cm}\textsc{SOLVE}$(\mathfrak{P_i})$
\STATE\hspace{1.0cm}\textsc{SOLVE}$(\mathfrak{P_j})$
\STATE\hspace{1.0cm}\textbf{return} $\mathfrak{P}_i$, $\mathfrak{P}_j$ as sub-problems
\STATE\hspace{0.5cm}\textbf{else}
\STATE\hspace{1.0cm}\textbf{return} $\mathfrak{P}$ as optimal solution 

\STATE
\STATE {\textsc{SOLVE}}$(\mathfrak{P})$
\STATE\hspace{0.5cm}$i \gets$ trajectory with newest conflict $C$
\STATE\hspace{0.5cm}\textbf{copy} $i$ twice: $j,k$
\STATE\hspace{0.5cm}\textbf{spatially resolve} $C$ for trajectory $j$
\STATE\hspace{0.5cm}\textbf{temporally plan} trajectory $j$
\STATE\hspace{0.5cm}\textbf{temporally plan} trajectory $k$
\STATE\hspace{0.5cm}\textbf{if} cost$(j)$ $<$ cost$(k)$ \textbf{then}
\STATE\hspace{1.0cm}\textbf{replace} $i$ with $j$ in $\mathfrak{P}$
\STATE\hspace{0.5cm}\textbf{else}
\STATE\hspace{1.0cm}\textbf{replace} $i$ with $k$ in $\mathfrak{P}$
\end{algorithmic}
\label{alg1}
\end{algorithm}

\subsection{Spatial Conflict Resolver}
The spatial conflict resolver component of OptWULF uses a very simple approach. Given the set of all previously and currently encountered spatial conflicts $\Psi_i$ at gate $i$, the lateral alignment $y_i$ at the gate is chosen which is the farthest to the right and which is conflict-free. Put mathematically, the proposed alignment $y_i$ must meet the following constraint:
\begin{equation}
\left ( \bigcup_{\psi \in \Psi_i}\psi \right ) \cap \left ( y_i-\frac{w_{veh}}{2} ~,~ y_i+\frac{w_{veh}}{2} \right ) = \emptyset
\label{eqn:optwulf-spatial-constraint}
\end{equation}

This new alignment is applied to gates $\{i\ldots n\}$. In other words, this alignment is applied at and downstream of the conflict. Any potentially newly introduced spatial conflicts are handled in their respective subsequent sub-problems.

\subsection{Temporal Planner}
Given a spatial path composed of $n$ gates spaced at distances of $l_i$ apart (i.e. $l_i = x_{i+1} - x_{i} ~ \forall i \in \{0 \ldots n-1\}$), the goal of the OptWULF temporal planner is to find a set of arrival times $\Vec{t} = [t_0 \ldots t_n ]$ at each respective gate which is kinematically feasible and results in the lowest possible travel time.

For each segment between adjacent gates, kinematic constraints for the maximum velocity of the vehicle $v_{f}$, maximum longitudinal acceleration $a_{lon,\max}$, and maximum lateral acceleration $a_{lat,\max}$ are applied. Both $v_{f}$ and $a_{lon,\max}$ are pre-defined constant attributes of the vehicle being routed. The maximum lateral acceleration is enforced for turning movements at the intersection by considering the turning radius $r_i$ and calculating the equivalent maximum velocity $v_{\max,i}=\sqrt{a_{lat,\max} / r_i}$.

The arrival time at each gate $i$ can be constrained to a given time window. Each time window constraint $\omega_i \in \Omega$ is a range $(\omega_{i,\min}~,~ \omega_{i,\max})$ during which the vehicle is allowed to cross gate $i$. The optimization problem of the temporal planner, which is described in detail below, is only capable of handling a single time window constraint per gate. However, in general, gates may have multiple free time windows to choose from. These represent, for example, the choice of whether to cross an intersection in front of or behind another vehicle. In order to deal with these more complicated constraint conditions, conflicts are converted to sets of time windows and feasible combinations of these windows are determined. The optimization problem is then solved in parallel for every one of these constraint combinations, and the best solution is accepted.

Since the high-level conflict-based search works by assigning conflict windows $\Phi_i$ for each gate $i$ but the temporal planner works with free time windows $\Omega_i$, the first step in finding feasible free time window constraint combinations is to use $\Phi_i$ to calculate $\Omega_i$ for each gate $i$. Intuitively, all continuous time ranges not blocked by a conflict are defined as free time windows. Stated mathematically, one is simply the complement of the other on the time range representing the remainder of the simulation:
\begin{equation}
    \Omega_i = \left [ t_{start}~,~ t_{end} \right ) \setminus \Phi_i
\end{equation}

\begin{figure*}[!t]
\centering
\includegraphics[width=181.35mm]{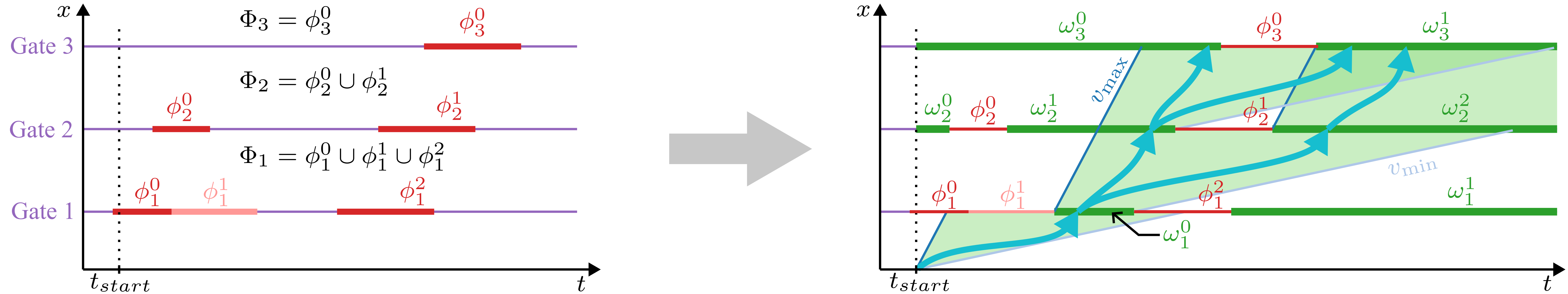}
\caption{Example conversion of temporal window constraints into free time windows and the determination of feasible free time window combinations.}
\label{fig:optwulf-feasible-window-combinations}
\end{figure*}

Feasible free time window combinations are determined by projecting each window $\omega_i$ downstream to the next gate with a window constraint as shown in Figure~\ref{fig:optwulf-feasible-window-combinations}. The beginning of the window is projected at a velocity of $v_{\max}$, and the end of the window is projected at a user-defined value $v_{\min}$. Setting $v_{\min}=0$ results in all kinematically feasible combinations being found, with the advantage that the planner exhaustively searches all possible options but with the disadvantage of significantly increased computation times. Experimentation showed that for the purposes of this study, a value of $v_{\min}=\unit[1]{m/s}$ provided a good balance between solution optimality and computation time. Mathematically, given a free time window constraint of $\omega_i^j=(\omega_{i,\min}^j~,~\omega_{i,\max}^j)$ at gate $i$, the feasible downstream free time windows at gate $m$ are:
\begin{align}
    \Omega_{m,feasible}^{i,j} = ~\Omega_m \cap \hat{\Omega}_m^{i,j} \\
    \hat{\Omega}_m^{i,j} = \left ( \omega_{i,\min}^j + \tau_{\max}(i,m) \,,~ \omega_{i,\max}^j + \tau_{\min}(i,m) \right ) \\
    \mathrm{given}~\Phi_\iota=\emptyset ~\forall \iota \in\{i+1 \ldots m-1\} \\
    \mathrm{where}~\tau_{\max}(i,m) = \sum_{\iota=i}^{m-1} \frac{l_\iota}{v_{\max,\iota}}\\
    \tau_{\min}(i,m) = \sum_{\iota=i}^{m-1} \frac{l_\iota}{v_{\min}}
\end{align}

For each set of feasible time window constraint combinations $\Omega$, the optimal kinematically feasible trajectory can be found by solving the following optimization problem, which we do using sequential least squares programming (SLSQP):

\begin{align}
    \min_{\Vec{t}} &&& t_n - t_0 \\
    \mathrm{s.t.} &&& t_0 = t_{start} \label{eqn:optwulf-opt-start-time-constraint} \\ 
    &&& t_i \geq t_{i-1} & \forall i \in \{1 \ldots n\} \label{eqn:optwulf-opt-nondecreasing} \\
    &&& v_i \leq \mathrm{\min}(v_{f}, v_{\max,i}) & \forall i \in \{0 \ldots n-1\} \label{eqn:optwulf-opt-velocity-constraint}\\
    &&& \left | a_i \right | \leq a_{\max} & \forall i \in \{0 \ldots n-2\} \label{eqn:optwulf-opt-accel-constraint} \\
    &&& t_i \geq \omega_{i,\min} & \forall \omega_i \in \Omega \label{eqn:optwulf-time-window-constraint-lower}\\
    &&& t_i \leq \omega_{i,\max} - \frac{l_{veh}}{v_i} & \forall \omega_i \in \Omega \label{eqn:optwulf-time-window-constraint-upper}
\end{align}

\noindent where
\begin{align}
    & v_i = \frac{l_i}{t_{i+1} - t_i} \\
    & a_i = \frac{v_{i+1} - v_{i}}{t_{i+1} - t_i}
\end{align}

The objective function, as mentioned previously, is simply the travel time of the vehicle (analog to the vehicle delay).
Equation~\ref{eqn:optwulf-opt-start-time-constraint} ensures that the the vehicle starts its trajectory at the required time (i.e.\ the registration time for newly arriving vehicles), Equation~\ref{eqn:optwulf-opt-nondecreasing} ensures that the gate crossing times are strictly non-decreasing (i.e.\ time travels forward), Equations~\ref{eqn:optwulf-opt-velocity-constraint}~\&~\ref{eqn:optwulf-opt-accel-constraint} enforce velocity and acceleration limits on the vehicle, and Equations~\ref{eqn:optwulf-time-window-constraint-lower}~\&~\ref{eqn:optwulf-time-window-constraint-upper} enforce the time window constraints.

If a feasible solution to this optimization problem is found, then the values of $\Vec{t}=[t_0 \ldots t_n]$ define the temporal component of the vehicle trajectory through the combination of time window constraints given by $\Omega$.
As mentioned, this procedure is applied to all feasible time window constraint combinations, and the one resulting in the lowest travel time is accepted as the solution by the temporal planner.

\section{Scenario Description}
\label{sec:scenario-description}
In this paper, fictitious four-way intersections of streets of width $w_{\mathrm{street}} \in \{6, 7, 8, 9, 10\}$ [m] and curb return radii of \unit[3]{m} are simulated using both FERSTT \cite{Malcolm.2025.FERSTT} and OptWULF.
For the sake of brevity, only scenarios with vehicle demand and no pedestrian or cyclist demand are considered.
As described in \cite{Malcolm.2025.FERSTT}, pedestrians and cyclists can be safely and efficiently incorporated into a lane-free AIM scheme by utilizing conventional traffic signal heads and employing an ``all-way green'' signal plan.
Such a signal plan consists of one phase during which AVs from and in all directions may access the intersection area (resolving their individual conflicts using the AIM controller) and another phase during which pedestrians and cyclists from and in all directions may access the intersection area (resolving their conflicts in an ad-hoc manner).
This concept is thus an extension of the ``simultaneous green'' concept already being successfully used in the Netherlands.
The AWG concept, combined with dedicated infrastructure for each of the main modes, ensures complete spatial and temporal separation of vehicular traffic streams from pedestrian and cyclist streams.
This provides increased safety and reduces the uncertainties involved in mixed traffic.
Our experiments have shown that for both FERSTT and OptWULF, given a sufficient green time per cycle for the AV movements, the saturation flow during the AV green phase is equal to that in scenarios without pedestrians or cyclists (and thus without any signal phases).
Therefore, the capacities of the vehicle-only scenarios given in this paper can in many cases be transferred to an arbitrary AWG signal plan by simply multiplying by the green time ratio.
In each scenario, vehicles register \unit[100]{m} upstream of the intersection and their trajectories are planned until \unit[100]{m} downstream of the intersection.
The gate spacing in OptWULF is \unit[10]{m} with the exception of the gate nearest the main intersection area, which placed at the end of the curb return and is spaced \unit[5]{m} from the next gate.
The speed limit in all scenarios is $\unit[30]{km/h}=\unit[8.3]{m/s}$. Vehicle dimensions are sampled from~\cite{Bock.2020.InD}, and a minimum lateral gap of \unit[0.1]{m} between vehicles is enforced (by increasing $w_{veh}$ for each vehicle by this amount). The minimum net time gap between vehicles is $g=\unit[1]{s}$. Each simulation consists of a warm-up time of \unit[60]{s} followed by a run time of \unit[600]{s}. All other simulation parameters are as described in~\cite{Malcolm.2025.FERSTT}.

Both symmetric and asymmetric demand patterns are tested. In the symmetric demand scenario, an equal flow of vehicles (with randomly distributed independent arrival times) comes from each direction, with 10\% turning left, 75\% continuing straight, and 15\% turning right. The nominal demand for the intersection is the sum of the demands of all the approaches. For example, for a demand of \unit[400]{veh/h}, each approach will have a demand of \unit[100]{veh/h}, of which \unit[10]{veh/h} will be left turners, and so on.
In asymmetric scenarios, an asymmetry factor $\beta$ is defined which determines how much of the demand on each axis comes from each direction. For example, if $\beta=0.5$, then the demand is symmetric, and if $\beta=1$, then there is only demand from the south and west approaches (and none from the north or east).

For each scenario, four randomly seeded simulations are performed for nominal demand levels in steps of \unit[400]{veh/h} until a mean AV delay of \unit[60]{s} exceeded. A best-fit curve is then calculated to fit the data.

Two benchmark scenarios are provided for context: a conventional lane-based intersection with traffic signal control, and the lane-based AIM SlotIIC~\cite{niels2022dissertation}. The parameters used for these benchmarks are described in detail in~\cite{Malcolm.2025.FERSTT}.

\section{Simulation Results}
\label{sec:simulation-results}
\begin{figure*}[!t]
\centering
\subfloat[]{\includegraphics[width=90mm]{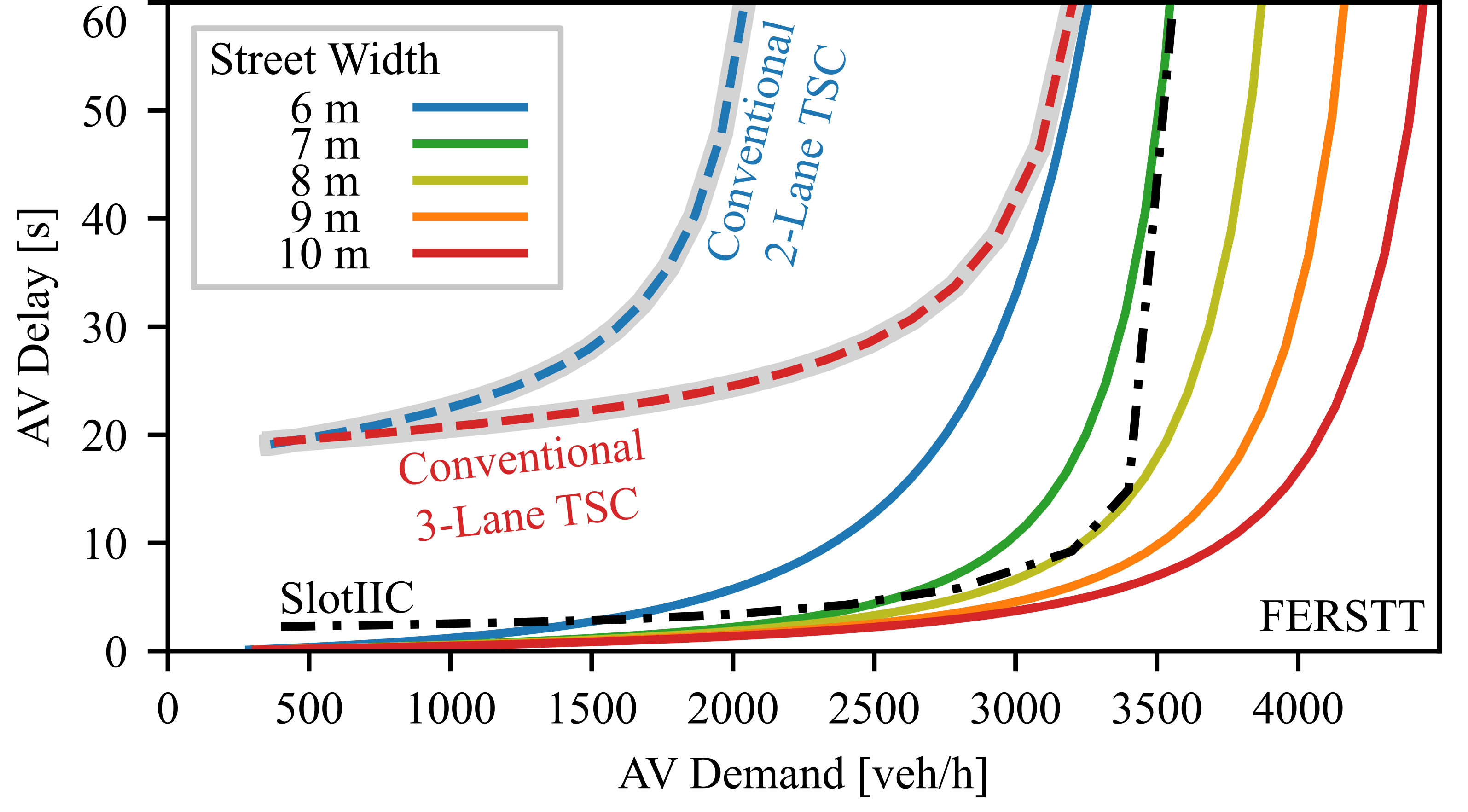}%
\label{fig:demand-delay-ferstt}}
\hfill
\subfloat[]{\includegraphics[width=90mm]{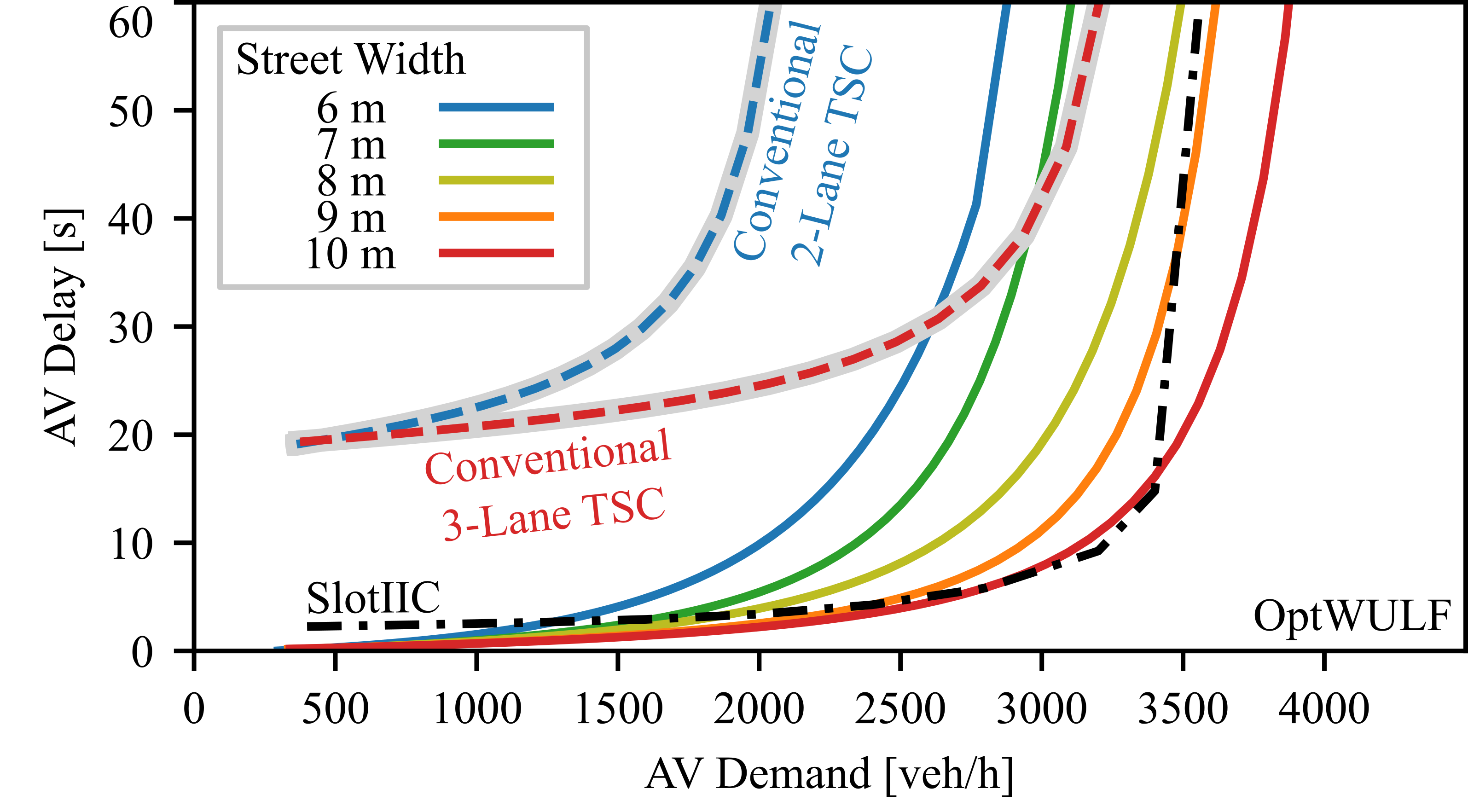}%
\label{fig:demand-delay-optwulf}}
\caption{Demand-delay relationship for (a) FERSTT and (b) OptWULF for several street widths. Also shown are benchmark results for conventional 2-lane and 3-lane traffic signal control (TSC) and the lane- based AIM SlotIIC.}
\label{fig:demand-delay}
\end{figure*}

The results of simulations of vehicle-only intersections for street widths of \unit[6--10]{m} for FERSTT and OptWULF are shown in Figure~\ref{fig:demand-delay} along with the results of the benchmarks.
As can be seen, despite OptWULF incorporating an optimization-based temporal planner component, it delivers a slightly lower capacity compared to FERSTT.
Both approaches are far more spatially efficient than the conventional lane-based traffic signal control (TSC), however. For example, FERSTT offers a similar capacity with \unit[6]{m} street widths to a conventional three-lane signalized intersection (with a street width of approx.\ \unit[9.6]{m}), a space savings of around 38\%. For the same capacity, OptWULF requires a street width of just over \unit[7]{m}, for a space savings of around 25\%.
At low and medium demands, the average AV delay in both FERSTT and OptWULF is extremely low, even less than that of the lane-based AIM SlotIIC.
FERSTT offers a similar capacity to SlotIIC for a similar street width, though the comparison is not entirely fair, since FERSTT employs a minimum net time gap of $g=\unit[1]{s}$ while SlotIIC uses $g=\unit[0.7]{s}$.
Taking this into consideration, it is clear that FERSTT is more spatially efficient (in terms of capacity per street width) than the lane-based SlotIIC.

The slightly lower capacity of OptWULF compared to FERSTT is likely due almost entirely to two factors.
Firstly, the spatial conflict resolver of OptWULF is relatively simplistic, which may result in some more complex spatial paths not being explored.
Secondly, while conflicts in the main intersection area are in some cases handled efficiently (taking into account the actual conflict region), some cases fall back to a less efficient approach that results in the intersection area being blocked for some vehicles longer than necessary.
Future work will address both of these shortcomings in an attempt to increase the capacity offered by OptWULF.

Based on the analytical results in Section~\ref{sec:analytical-saturation-flow}, one would expect to see a similar nonlinear relationship between capacity and street width in the lane-free scenarios.
While this effect cannot be clearly seen for FERSTT, it is clear that in OptWULF, the capacity increases more for some street width steps (\unit[7--8]{m} and \unit[9--10]{}m) than for others (\unit[6--7]{m} and \unit[8--9]{m}).

\begin{figure*}[!hbt]
\centering
\subfloat[]{\includegraphics[width=79mm]{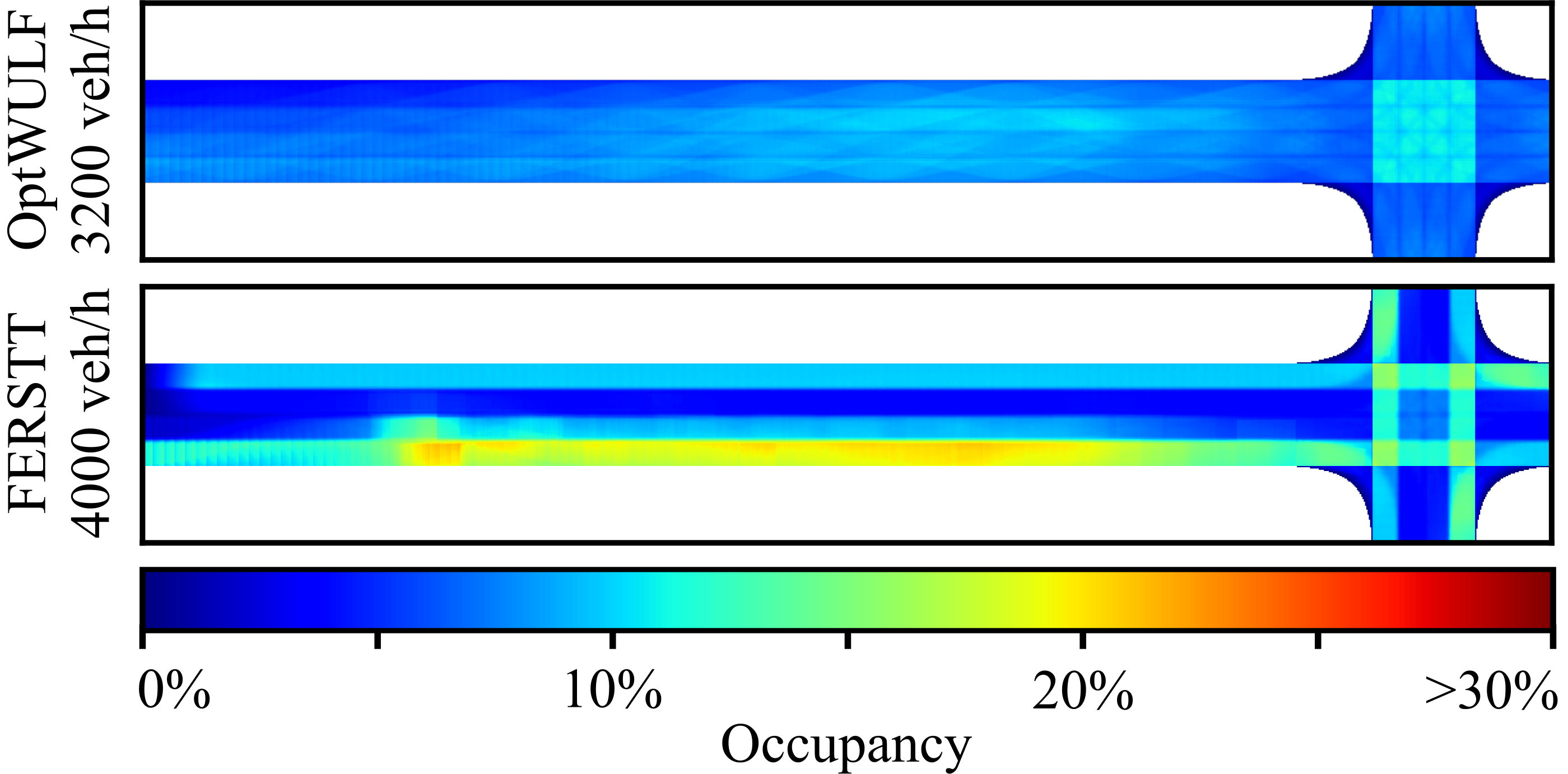}%
\label{fig:heatmaps-occupancy}}
\hfill
\subfloat[]{\includegraphics[width=79mm]{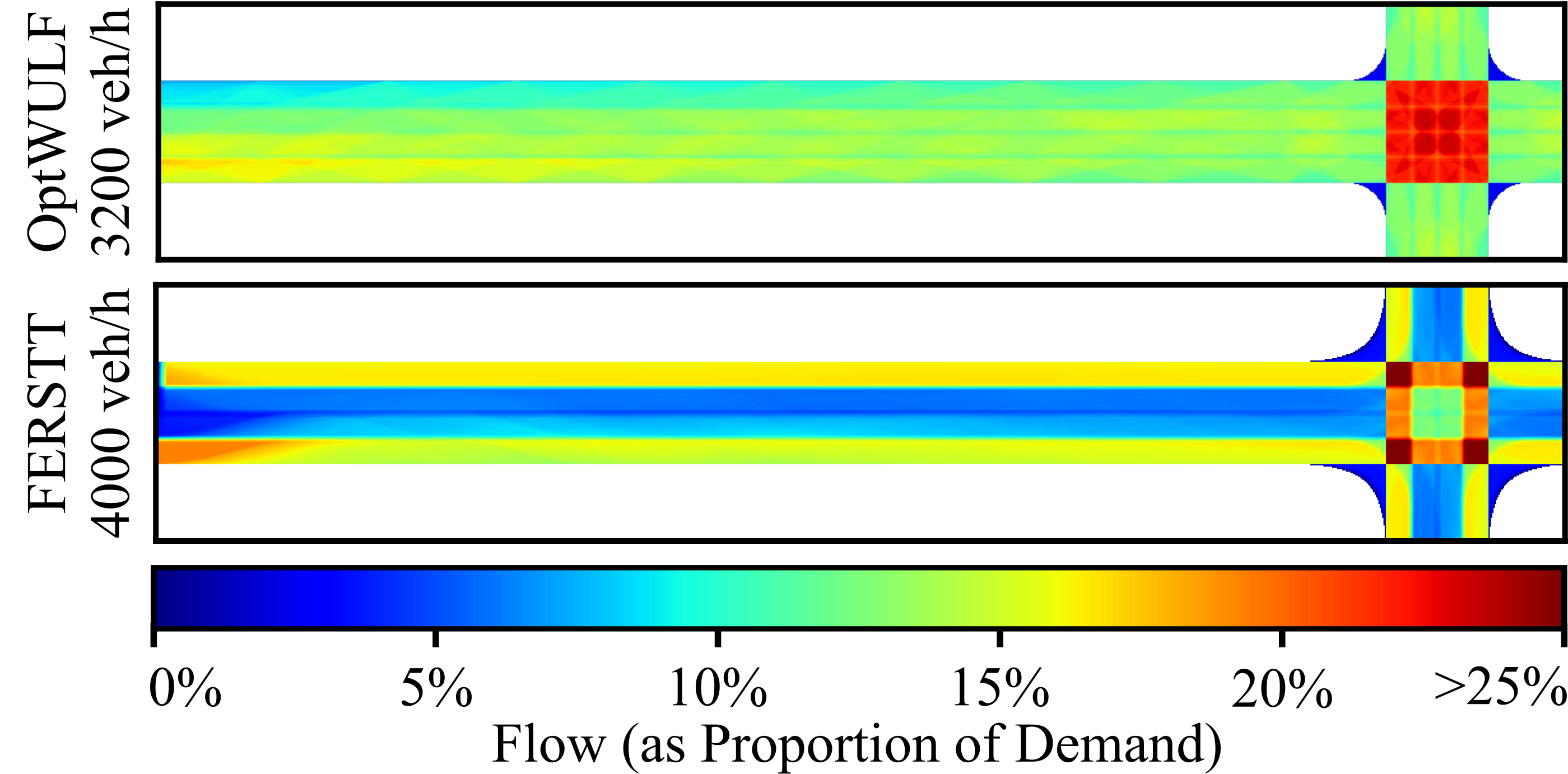}%
\label{fig:heatmaps-flow}}
\caption{Heatmaps of (a) occupancy and (b) flow for both OptWULF and FERSTT with a street width of \unit[8]{m} at high (near-capacity) demand.}
\label{fig:heatmaps}
\end{figure*}

The most interesting difference between FERSTT and OptWULF is that while FERSTT employs a traffic-shaping penalty term in its search heuristic that encourages vehicles to ``keep right'' except when disadvantageous to do so, OptWULF allows vehicles to choose any spatial alignment freely. This includes allowing contraflow (i.e.,\ left-side driving), which can be helpful in reducing delays for left-turning vehicles, for example.
OptWULF therefore also utilizes the available street and intersection space more evenly. This difference can be clearly seen in Figure~\ref{fig:heatmaps}, which shows the occupancy and flow for each cell on a \unit[10]{cm} grid for the \unit[8]{m} street width scenario at high (near-capacity) demand.

In the occupancy plots (Figure~\ref{fig:heatmaps-occupancy}), the effect of the explicitly-defined dawdling behavior in FERSTT can be seen in the higher occupancy a short distance upstream of the intersection.
OptWULF also appears to exhibit some spontaneously emerging dawdling behavior as well, as the occupancy in the same region is slightly higher (though the difference is not as pronounced).
In the flow plots (Figure~\ref{fig:heatmaps-flow}), the effect of the keep-right penalty of FERSTT is clearly visible. The middle of the street and intersection area are not nearly as highly utilized as the edges. With OptWULF, however, nearly every part of the street and intersection area is equally utilized.

\begin{figure}[!bt]
\centering
\includegraphics[width=88.566mm]{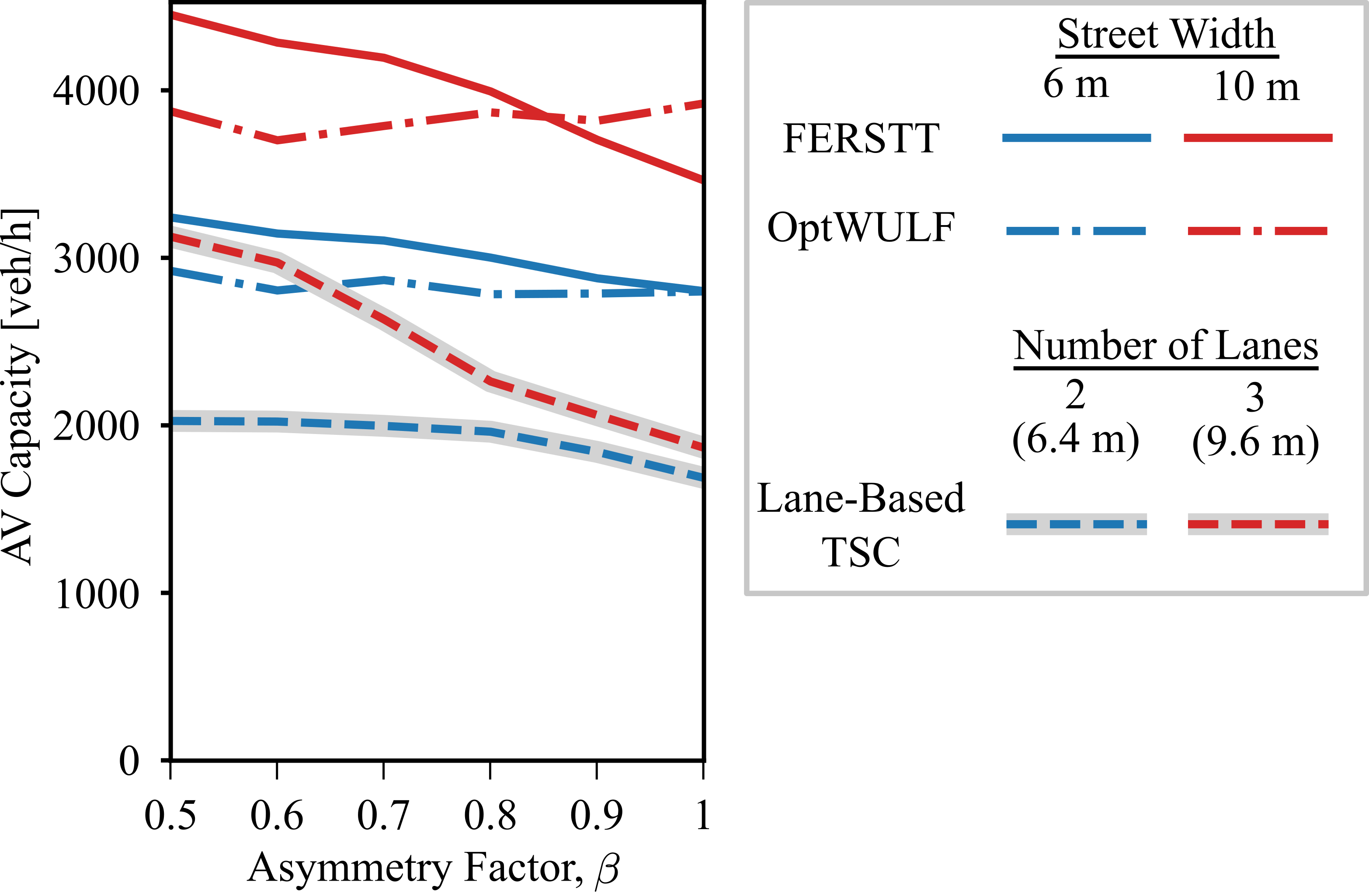}
\caption{Capacity of OptWULF and FERSTT under various degrees of demand asymmetry $\beta$ for \unit[6]{m} and \unit[10]{m} street widths. Also shown are the results of the baseline two-~and three-lane lane-based signalized intersection. For the purposes of this figure, capacity is defined as the demand at which the delay is equal to \unit[55]{s} (based on the fitted demand-delay curve).}
\label{fig:asymmetric-demand}
\end{figure}

One of the most practical benefits of lane-free traffic is that it allows for a dynamic and spontaneous reaction to asymmetric demand patterns.
To illustrate this, both OptWULF and FERSTT were used to simulate demand patterns with varying asymmetry factors $\beta$ (as defined in Section~\ref{sec:scenario-description}).
The results of these simulations are shown in Figure~\ref{fig:asymmetric-demand}.

Here it can be seen that for the narrow street scenarios (\unit[6]{m} for lane-free scenarios and two lanes for lane-based scenarios), asymmetric demands are handled rather well, with only slight decreases in capacity.
For the 2-lane lane-based TSC, the capacity remains nearly constant until a demand asymmetry factor of $\beta=0.8$ is reached, after which the capacity slightly decreases.
The reason for this is that left-turning vehicles under high symmetric demand must wait longer for a sufficient gap in the oncoming traffic stream, increasing delay and reducing the capacity of their lane.
As asymmetry increases, the left-turning vehicles in the major stream do not have to wait as long, increasing that lane's capacity, but this effect is canceled out by the lesser utilization of the minor direction lane.
Under highly asymmetric demand, the effect of the lack of utilization of the minor direction lane becomes larger, reducing the overall capacity.
For FERSTT with a \unit[6]{m} street width, the reduction in capacity is relatively constant with respect to $\beta$.
For OptWULF with a \unit[6]{m} street width, the capacity is only very slightly lower under fully asymmetric demand compared to symmetric demand.

For the wider street scenarios (\unit[10]{m}~/~three lanes), the effect of asymmetry is much more pronounced.
At $\beta=1$ (fully asymmetric), the capacity of the three-lane lane-based scenario is only slightly better than that of the two-lane scenario.
FERSTT is also sensitive to asymmetric demand, though to a slightly lesser extent than the lane-based signalized control.
The capacity of OptWULF, on the other hand, is nearly constant across all demand asymmetry levels.
This makes sense, as the lack of a keep-right heuristic (as in FERSTT) allows for left-turning vehicles to resolve their conflicts far more flexibly, meaning they do not have as much of an effect on subsequent vehicles.

\section{Conclusion}
In this paper, the increased spatial efficiency of future lane-free traffic was investigated. For lane-free roadways, a simple analytical approach was used and it was found that the relationship between capacity and street width is continuous but nonlinear under the assumption of current vehicle dimensions.
If narrow vehicles become more popular, then this relationship can become nearly linear.
To determine the capacity and properties of lane-free AIM, two AIM approaches were tested: FERSTT \cite{Malcolm.2025.FERSTT} and a novel approach introduced in this paper, OptWULF.
While OptWULF does not beat FERSTT in capacity, it does exhibit interesting properties, such as an even utilization of the entire drivable area and the ability to handle symmetric and asymmetric demand patterns equally well.

For transportation planners and engineers in a potential future lane-free world, the opportunity exists to use the increased spatial efficiency of lane-free traffic not only to increase capacity and reduce delays for AV passengers, but to reduce the space dedicated to AV infrastructure and rededicate it to improved pedestrian and cyclist facilities and/or green space.
The continuous relationship between street width and capacity in particular means that planners in such a future will have much finer-grained control over the balance of level of service across all transport modes.

\section*{Acknowledgments}
Portions of this work are part of the project ``Simulation and Organization of Future Lane-Free Traffic'' funded by the German Research Foundation (DFG), under the project number BO~5959/1-1.



\bibliographystyle{IEEEtran}
\bibliography{references.bib}

\newpage

 


%
\begin{IEEEbiography}[{\includegraphics[width=1in,height=1.25in,clip,keepaspectratio]{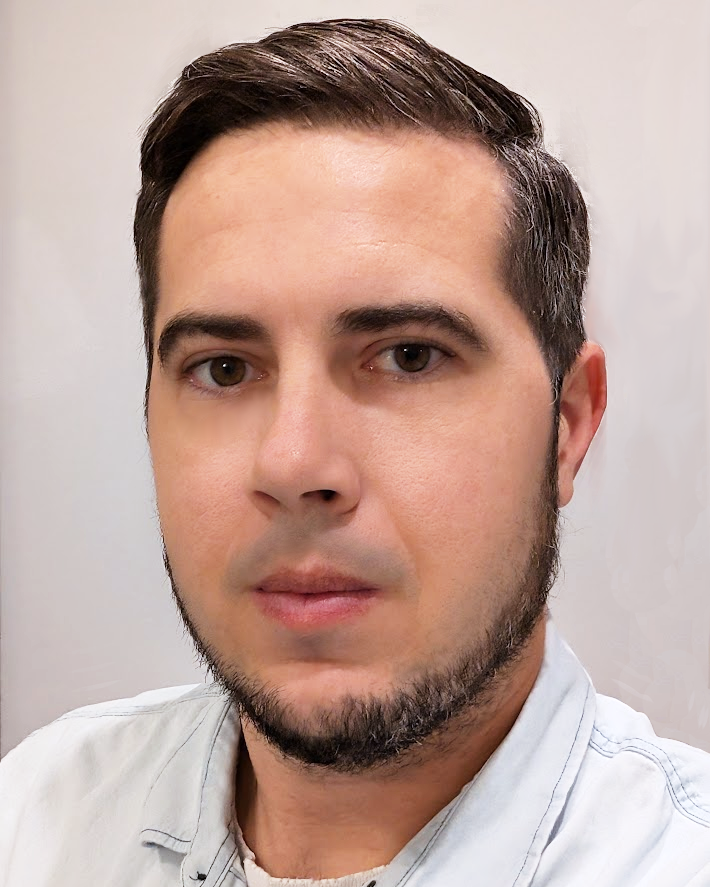}}]{Patrick Malcolm}
received the M.Sc.\ degree in Transportation Systems from the Technical University of Munich. He is currently working and pursuing a Ph.D.\ at the Chair of Traffic Engineering and Control at the Technical University of Munich. His main research interests include lane-free traffic, pedestrian modeling, and microscopic traffic simulation.
\end{IEEEbiography}

\begin{IEEEbiography}[{\includegraphics[width=1in,height=1.25in,clip,keepaspectratio]{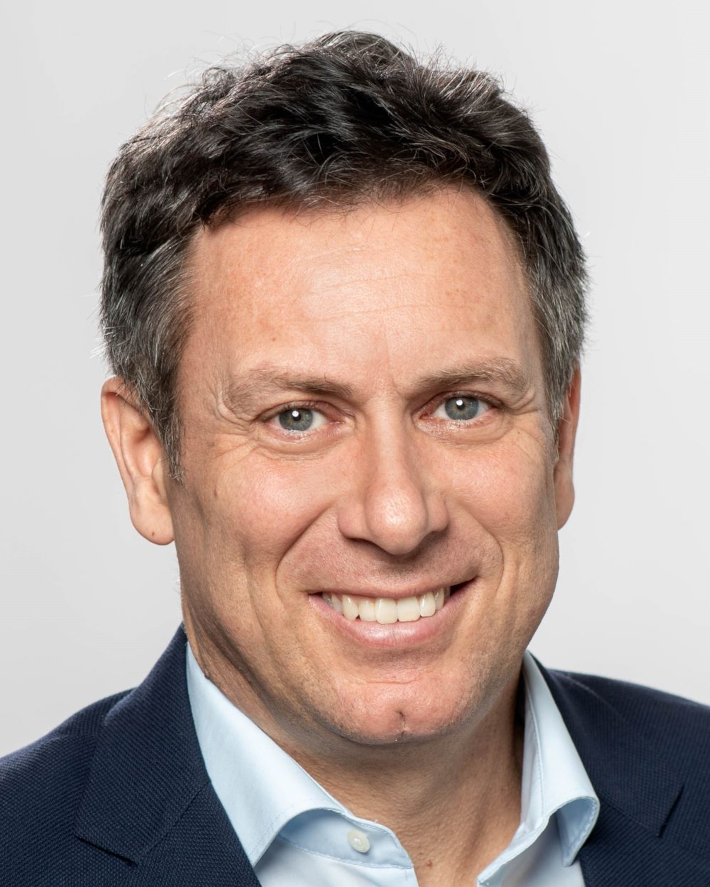}}]{Klaus Bogenberger}
received the Diploma degree in civil engineering and the Ph.D.\ degree in traffic engineering from the Technical University of Munich, Germany, in 1996 and 2001, respectively.
He was a Research Engineer with BMW Group from 2001 to 2008. At first, he was responsible for the traffic flow theory and models with the Department of Science and Transportation and later on, he was
with the Department of Corporate Quality. From 2008 to 2011, he was the Managing Director and a Partner of TRANSVER GmbH (consultant office
for transport planning and traffic engineering), Munich and Hannover. From 2012 to 2019, he was with Bundeswehr University Munich, as a Professor of traffic engineering. Since 2020, he has been the Chair of Traffic Engineering and Control and the head of the Mobility System Engineering Department, Technical University of Munich. His main research interests include on-demand systems, tradeable travel credits, and autonomous vehicles.
\end{IEEEbiography}

\vfill

\end{document}